\documentclass[prc,superscriptaddress,preprintnumbers,showpacs,nofootinbib,a4paper,twocolumn]{revtex4}

\usepackage{verbatim}
\usepackage[latin1]{inputenc}
\usepackage{graphicx,amssymb} 

\usepackage{dcolumn}
\usepackage{color}%

\graphicspath{{./Figures-EPS/}{../Figures-EPS/}}
\newcommand{\mc}[3]{\multicolumn{#1}{#2}{#3}}
\newcommand{\fnm}[1]{\footnotemark[#1]}
\newcommand{\fzd}{\affiliation{Forschungszentrum Dresden-Rossendorf (FZD), Dresden, Germany}}

\newcommand{\pd}{\affiliation{Istituto Nazionale di Fisica Nucleare (INFN), Sezione di Padova, Italy}}

\begin{document}

\title{Resonance strengths in the $^{14}$N(p,$\gamma$)$^{15}$O and $^{15}$N(p,$\alpha\gamma$)$^{12}$C reactions} 

\date{As accepted by Phys.Rev. C, 28 May 2010}
\thispagestyle{empty}

\author{Michele Marta}\fzd
\author{Erik Trompler}\fzd
\author{Daniel Bemmerer}\fzd
\author{Roland Beyer}\fzd
\author{Carlo Broggini}\pd
\author{Antonio Caciolli}\pd
\author{Martin Erhard}\pd
\author{Zsolt F\"ul\"op}\affiliation{Institute of Nuclear Research (ATOMKI), Debrecen, Hungary}
\author{Eckart Grosse}\fzd
\author{Gy\"orgy Gy\"urky}\affiliation{Institute of Nuclear Research (ATOMKI), Debrecen, Hungary}
\author{Roland Hannaske}\fzd
\author{Arnd R. Junghans}\fzd
\author{Roberto Menegazzo}\pd
\author{Chithra Nair}\fzd
\author{Ronald Schwengner}\fzd
\author{Tam\'as Sz\"ucs}\affiliation{Institute of Nuclear Research (ATOMKI), Debrecen, Hungary}\fzd
\author{Simone Vezz\'u}\affiliation{Coordinamento Interuniversitario Veneto per le Nanotechnologie (CIVEN), Venice, Italy}
\author{Andreas Wagner}\fzd
\author{Dmitry Yakorev}\fzd

\begin{abstract}
The $^{14}$N(p,$\gamma$)$^{15}$O reaction is the slowest reaction of the carbon-nitrogen-oxygen cycle of hydrogen burning in stars. As a consequence, it determines the rate of the cycle. The $^{15}$N(p,$\alpha\gamma$)$^{12}$C reaction is frequently used in inverse kinematics for hydrogen depth profiling in materials. 
The $^{14}$N(p,$\gamma$)$^{15}$O and $^{15}$N(p,$\alpha\gamma$)$^{12}$C reactions have been studied simultaneously, using titanium nitride targets of natural isotopic composition and a proton beam.
The strengths of the resonances at  $E_{\rm p}$ = 1058\,keV in $^{14}$N(p,$\gamma$)$^{15}$O and at $E_{\rm p}$ = 897 and 430\,keV in $^{15}$N(p,$\alpha\gamma$)$^{12}$C have been determined with improved precision, relative to the well-known resonance at $E_{\rm p}$ = 278\,keV in $^{14}$N(p,$\gamma$)$^{15}$O.
The new recommended values are $\omega\gamma$ = 0.353$\pm$0.018, 362$\pm$20, and 21.9$\pm$1.0\,eV for their respective strengths.
In addition, the branching ratios for the decay of the $E_{\rm p}$ = 1058\,keV resonance in $^{14}$N(p,$\gamma$)$^{15}$O have been redetermined.
The data reported here should facilitate future studies of off-resonant capture in the $^{14}$N(p,$\gamma$)$^{15}$O reaction that are needed for an improved R-matrix extrapolation of the cross section. In addition, the data on the 430\,keV resonance in $^{15}$N(p,$\alpha\gamma$)$^{12}$C may be useful for hydrogen depth profiling.

\end{abstract}



\maketitle

\section{Introduction}
The carbon-nitrogen-oxygen (CNO) cycle \cite{Weizsaecker38-PZ,Bethe39-PR_letter} dominates stellar hydrogen burning for temperatures of 20-150\,MK \cite{Iliadis07-Book}. A quantitative understanding of its rate affects, for instance, the dredge-up of nucleosynthetic material to the surface of so-called carbon stars \cite[]{Herwig06-PRC}. At lower temperatures, hydrogen burning is dominated by the proton-proton (pp) chain instead. In our Sun, the CNO cycle accounts for just 0.8\% of energy production \cite{Bahcall04-PRL}, but it provides an interesting neutrino signal. 

The solar CNO neutrino flux is proportional to the abundance of carbon and nitrogen in the solar core \cite{Haxton08-ApJ}. This abundance is closely connected to the so-called solar composition problem: There are newly revised elemental abundance data for the solar atmosphere from an improved analysis of Fraunhofer absorption lines \cite{Asplund09-ARAA}. This new elemental composition, when fed into the accepted standard solar model, leads to predicted 
observables such as the sound speed and density profiles, the depth of the convective zone, and the abundance of helium on the surface \cite{Turck04-PRL,Bahcall06-ApJS,Serenelli09-ApJL}, that are in disagreement with helioseismological data \cite{Basu09-ApJ}. The solar composition problem might be solved if the elemental composition is different in the solar core than in the atmosphere.

Two key ingredients for a study of the carbon and nitrogen abundance in the solar core are already available: First, the experimental data on the flux of $^8$B neutrinos from the Sun have reached a precision of 3\% for the Super-Kamiokande I data \cite{Hosaka06-PRD}, and the oscillation parameters for solar neutrinos have by now been well-constrained, most notably by data from the SNO \cite{SNO08-PRL} and KamLAND \cite{KamLAND08-PRL} neutrino detectors. The flux of solar $^7$Be neutrinos is under study in the Borexino detector and currently known with 10\% precision \cite{Borexino08-PRL}, a number that is expected to improve in the near future. Second, the nuclear reaction cross sections involved in producing these neutrinos are rather well-known \cite{NaraSingh04-PRL,Bemmerer06-PRL,Brown07-PRC,DiLeva09-PRL,Junghans10-PRC}. Therefore, the $^8$B and $^7$Be neutrinos can be used as a thermometer \cite{Haxton08-ApJ} to measure the temperature of the solar core (approximately 16\,MK). 

A third ingredient, the flux of CNO neutrinos from the $\beta^+$ decay of $^{13}$N and $^{15}$O, has not yet been measured online. However, it is believed that both Borexino and the planned SNO+ detector \cite{Chen05-NPBPS} can provide such data in the near future. A fourth ingredient are the nuclear reaction rates involved in the production of the CNO neutrinos. The rate of the reaction controlling the rate, $^{14}$N(p,$\gamma$)$^{15}$O, is currently known with only 8\% precision \cite{Marta08-PRC}, not enough to resolve the solar composition problem.

\begin{figure*}[thb]
  \includegraphics[angle=0,width=0.33\textwidth]{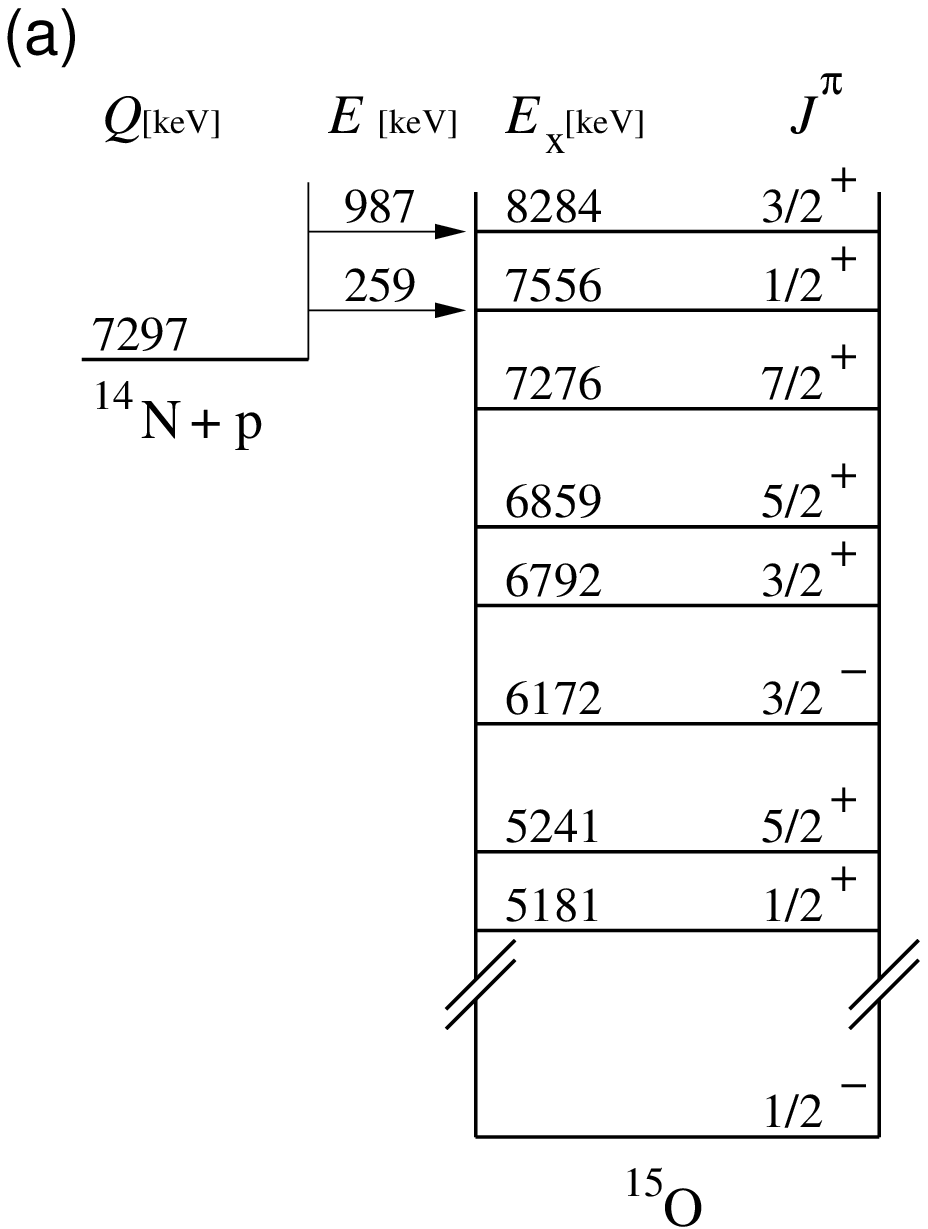}
  \hspace{0.09\textwidth}%
  \includegraphics[angle=0,width=0.54\textwidth]{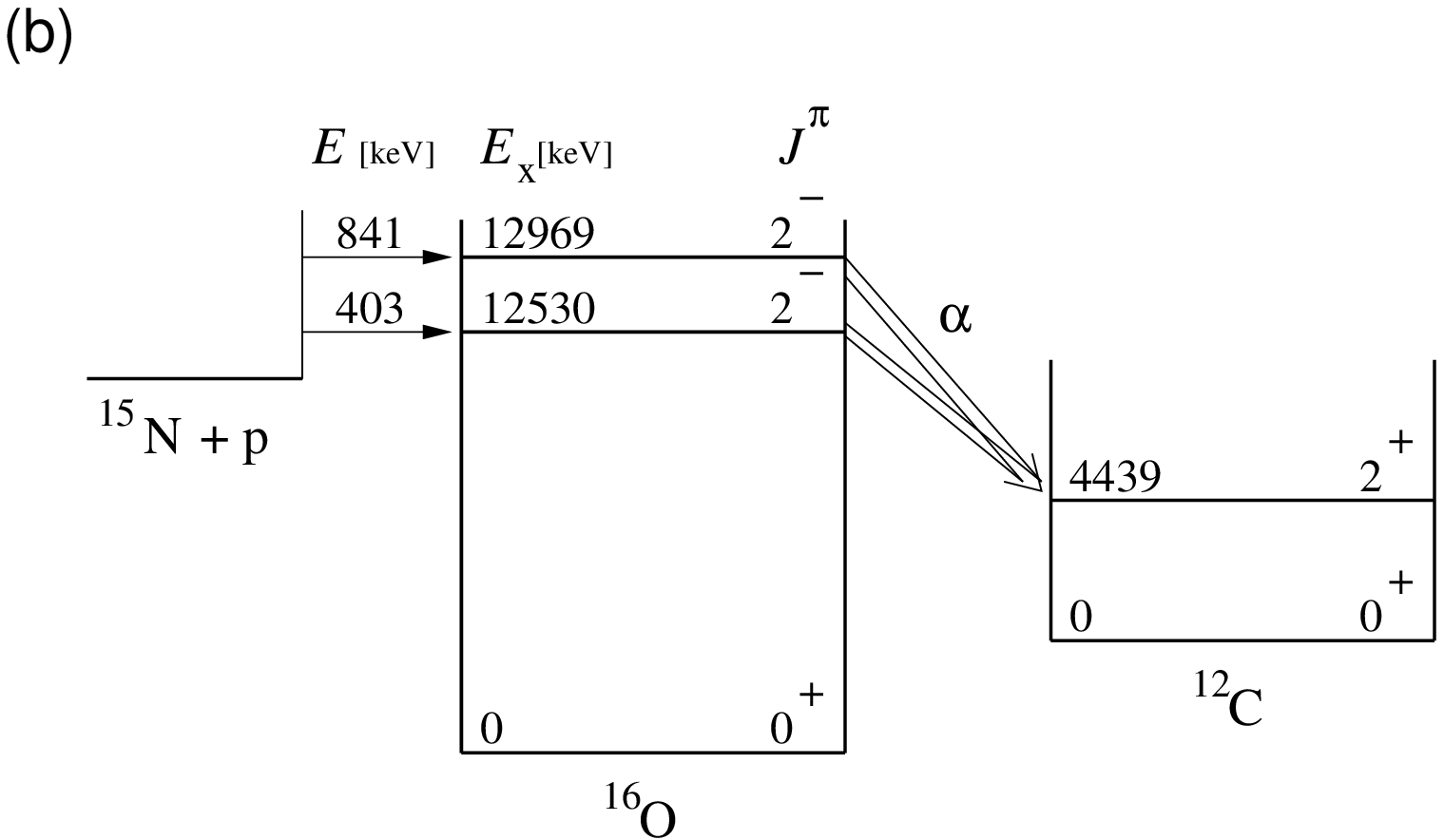}
\caption{(a) Level scheme of $^{15}$O showing the first eight excited states (energies $E_{\rm x}$ and $Q$-value $Q$ \cite{Ajzenberg13-15,Imbriani05-EPJA} in keV) relevant to the $^{14}$N(p,$\gamma$)$^{15}$O reaction. (b) Level scheme of the $^{15}$N(p,$\alpha\gamma$)$^{12}$C reaction ($Q$-value 4965\,keV). 
}
\label{fig:levelscheme}
\end{figure*}

The $^{14}$N(p,$\gamma$)$^{15}$O reaction proceeds through capture to a number of excited states and the ground state of $^{15}$O (fig.~\ref{fig:levelscheme}, left panel). The last comprehensive study of this reaction covering a wide energy range goes back to the 1980's \cite{Schroeder87-NPA}. 
In more recent years, many of the results of Ref.~\cite{Schroeder87-NPA} have come under renewed scrutiny. The $\gamma$-width of the subthreshold state at 6792\,keV is now believed to be much lower than assumed in Ref.~\cite{Schroeder87-NPA}. This conclusion was reached in Doppler shift attenuation experiments \cite{Bertone01-PRL,Schuermann08-PRC}, a Coulomb excitation study \cite{Yamada04-PLB}, and R-matrix fits \cite{Angulo01-NPA,Mukhamedzhanov03-PRC,Marta08-PRC}. The off-resonant capture cross-sections have also been re-investigated at energies\footnote{In the following text, $E_{\rm p}$ denotes the proton energy in the laboratory system, $E$ the center of mass energy.} 70\,keV $<$ $E$ $<$ 500\,keV, in some cases significantly revising the Ref.~\cite{Schroeder87-NPA} data \cite{Formicola04-PLB,Imbriani05-EPJA,Runkle05-PRL,Lemut06-PLB,Bemmerer06-NPA,Marta08-PRC}. An analyzing power study even questioned the transition mode for some decays of excited states \cite{Nelson03-PRC}. In summary, the new recommended total cross section at astrophysical energies \cite{Marta08-PRC} is a factor two lower than previously believed \cite{Schroeder87-NPA}, so the accepted reaction rate databases for astrophysical modeling \cite{CF88-ADNDT,NACRE99-NPA,Adelberger98-RMP} will have to be revised accordingly.

Despite all the efforts on the $\gamma$-width of the 6792\,keV state and on low-energy cross sections, for higher energies $E$ $\geq$ 500\,keV
no experimental re-investigation of the $^{14}$N(p,$\gamma$)$^{15}$O cross section has been performed since the 1980's. However, for this reaction also precise high-energy data play a role \cite{Angulo01-NPA,Mukhamedzhanov03-PRC} in extrapolating the cross section in the R-matrix framework to ultra-low astrophysical energies such as the solar Gamow peak at 28\,keV. 

The logical first step of a re-investigation of $^{14}$N(p,$\gamma$)$^{15}$O at $E$ $\geq$ 500\,keV is a renewed study of the sharp resonance at $E_{\rm p}$ = 1058\,keV. Due to the complicated R-matrix scheme with at least five poles and also direct capture contributions, its parameters cannot directly be transformed into formal R-matrix parameters. However, they can be used as outside constraints for an R-matrix fit, and as normalization points for off-resonant capture studies. The most precise available reference point for a study of this high-energy resonance is the low-energy $^{14}$N(p,$\gamma$)$^{15}$O resonance at $E_{\rm p}$ = 278\,keV. Its resonance strength 
\begin{equation} 
\omega\gamma = \frac{2J+1}{(2j_1+1)(2j_2+1)} \frac{\Gamma_{1} \Gamma_{2}}{\Gamma}
\end{equation}
(with $j_1, j_2, J$ the total angular momenta and $\Gamma_i$ the widths) has been measured several times with consistent results \cite{Becker82-ZPA,Runkle05-PRL,Imbriani05-EPJA,Lemut06-PLB}, and based on these works an averaged value of $\omega\gamma_{278}$ = 13.1$\pm$0.6\,meV has recently been recommended \cite{Seattle09-workshop}. The resonance is very narrow \cite{Borowski09-AIPCP}, and the isotropy of its emitted $\gamma$-rays makes it also a convenient tool for a relative $\gamma$-efficiency calibration.

Two further reference points offer themselves, the resonances at $E_{\rm p}$ = 430 and 897\,keV in the $^{15}$N(p,$\alpha\gamma$)$^{12}$C reaction. For practical reasons, many $^{14}$N targets contain also $^{15}$N with its natural and exceptionally stable isotopic abundance of 0.3663\%. The two $^{15}$N(p,$\alpha\gamma$)$^{12}$C resonances are rather sharp and sufficiently strong to stand out despite the small isotopic abundance of $^{15}$N. 
The resonance at $E_{\rm p}$ = 430\,keV is frequently used for hydrogen depth profiling using 6.39\,MeV $^{15}$N ions \cite[e.g.]{Wielunski02-NIMB}, with the 4.439\,MeV $\gamma$-ray from the reaction being detected. Owing to this application, the total energetic width $\Gamma$ of this resonance has been studied frequently \cite[]{Maurel83-NIM,Zinke85-ZPA,Osipowicz87-NIMB}. However, its  $\omega\gamma$ has so far been measured only once with precision better than 10\% \cite{Becker95-ZPA}. 

The aim of the present work is to provide precise values for the strengths of three resonances: The resonance at $E_{\rm p}$ = 1058\,keV in $^{14}$N(p,$\gamma$)$^{15}$O and the resonances at $E_{\rm p}$ = 430 and 897\,keV in $^{15}$N(p,$\alpha\gamma$)$^{12}$C. In addition, the branching ratios of the decay of the $E_{\rm p}$ = 1058\,keV resonance in $^{14}$N(p,$\gamma$)$^{15}$O are re-studied.
These three resonances may then serve as normalization points in a re-investigation of the $^{14}$N(p,$\gamma$)$^{15}$O reaction for $E$ $>$ 500\,keV. In addition, improved absolute strength values for the $^{15}$N(p,$\alpha\gamma$)$^{12}$C resonances will aid an absolute calibration of hydrogen depth profiling with $^{15}$N beams.

\section{Experimental setup}

\subsection{Ion beam, beam transport, and target chamber}

\begin{figure}[tbh]
  \includegraphics[angle=0,width=\columnwidth]{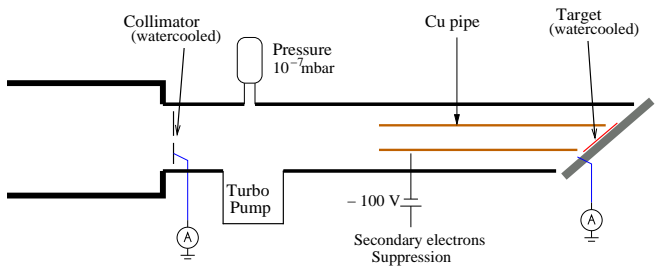}	
\caption{(Color online) Schematic view of the target chamber (see text).}
\label{fig:chamber}
\end{figure}

The H$^+$ beam for the experiment was provided by the 3\,MV Tandetron accelerator~\cite{Friedrich96-NIMA} at Forschungszentrum Dresden-Rossendorf (FZD).
The beam reached the target chamber (fig.~\ref{fig:chamber}) after passing a switching magnet, an electrostatic quadrupole lens, electrostatic dipoles and a neutral particle trap. The neutral particle trap consisted of an electric dipole positioned 1\,m upstream from the target, bending the beam by 7$^\circ$. The neutral particles continued at 0$^\circ$ and were absorbed on 
the internal wall.

A copper collimator of 5\,mm diameter was placed 45\,cm upstream from the target. A 12\,cm long copper pipe of 2\,cm diameter was inserted coaxial to the beam, at 5\,mm distance from the target. The copper pipe was biased with -100\,V to suppress secondary electrons from the target which might affect the electrical beam current reading. It is estimated that the electrical currents are accurate to $\pm$1.0\% in this Faraday cup. The vacuum measured at 40\,cm distance from the target was typically 1$\cdot$10$^{-7}$\,mbar during the irradiations.

The beam intensity on the target ranged from 1-15\,$\mu$A. The current on the collimator was always comparable in size to the target current, so no beam wobbling was necessary.
The absolute proton beam energy $E_{\rm p}$ was calibrated based on the known energies of eight resonances in the $^{14}$N(p,$\gamma$)$^{15}$O, $^{15}$N(p,$\alpha\gamma$)$^{12}$C, and $^{27}$Al(p,$\gamma$)$^{28}$Si reactions ranging in energy from $E_{\rm p}$ = 278 to 2047\,keV. The observed beam energy spread was 1.1\,keV (FWHM) at $E_{\rm p}$ = 897\,keV. 

\subsection{Targets}
\label{sec:target}

\begin{figure}[tbh]
  \includegraphics[angle=-90,width=.8\columnwidth]{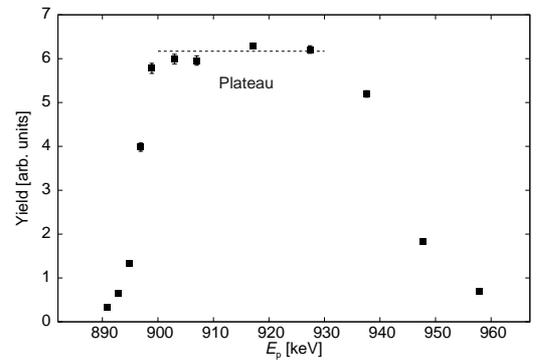} 
\caption{Depth profile of nitrogen content in the target, obtained by scanning the $^{15}$N(p,$\alpha\gamma$)$^{12}$C resonance at $E_{\rm p}$ = 897\,keV for a virgin target. Plotted is the yield of the 4.44\,MeV $\gamma$-ray versus proton energy (squares) and average yield in the plateau region (dashed line).}
\label{fig:target}
\end{figure}

For the experiment, titanium nitride targets have been used. They were produced with the reactive sputtering technique at the CIVEN facility in Venice/Italy, using nitrogen gas of natural isotopic abundance. This technique usually leads to highly stable targets with stoichiometry close to Ti$_1$N$_1$. 
The abundance of $^{15}$N in the nitrogen contained in atmospheric air, (0.3663$\pm$0.0004)\% \cite{Coplen02-PAC}, has been found to be exceedingly stable \cite{Mariotti83-Nature}, so it is even defined as the abundance standard by the International Union of Pure and Applied Chemistry \cite{Coplen02-PAC}. 
In a recent study using commercial nitrogen tank gas of natural abundance, the $^{15}$N/$^{14}$N ratio was checked by mass spectrometry and found to be consistent with the natural abundance \cite{Bemmerer09-JPG}. 
For the purpose of the present work, the standard isotopic abundance \cite{Coplen02-PAC} is assumed to hold with 1.0\% uncertainty \cite{Junk58-GCA}. Any effects of target degradation under the ion beam are expected to derive from atomic processes with negligible isotopic effects, so it is assumed here that the relevant behavior of the $^{15}$N atoms tracks that of the $^{14}$N atoms. Consequently, the same targets could be used for a parallel study of proton capture on $^{14}$N and $^{15}$N.

Four different samples have been used, all consisting of a 200\,$\mu$g$/$cm$^2$ thick layer of TiN on a 0.22\,mm thick tantalum backing. The targets were placed tilted by 55$^{\circ}$ with respect to the beam axis and were directly watercooled. 

The nitrogen content of the targets and its distribution have been checked at regular intervals by scanning the $^{15}$N(p,$\alpha\gamma$)$^{12}$C resonance at $E_{\rm p}$ = 897\,keV (width $\Gamma_{\rm lab}$ = 1.57\,keV \cite{Tilley_16-17}, slightly larger than the observed beam energy spread), recording the yield of the 4.44\,MeV $\gamma$-ray from the decay of the first excited state of $^{12}$C. The targets showed a rectangular depth profile (fig. \ref{fig:target}), with an energetic width of typically 50\,keV at $E_{\rm p}$ = 897\,keV and at 55$^\circ$. The observed high-energy tail of the target is consistent with the expected 13\,keV energy straggling at the target end. The plateau of this resonance scan was allowed to decrease by up to 15\% under irradiation, then the target was replaced.

\subsection{Detection of emitted photons}
\begin{figure*}[tbh]
	\includegraphics[angle=0,width=0.4\textwidth,trim=10cm 0 0 0,clip]{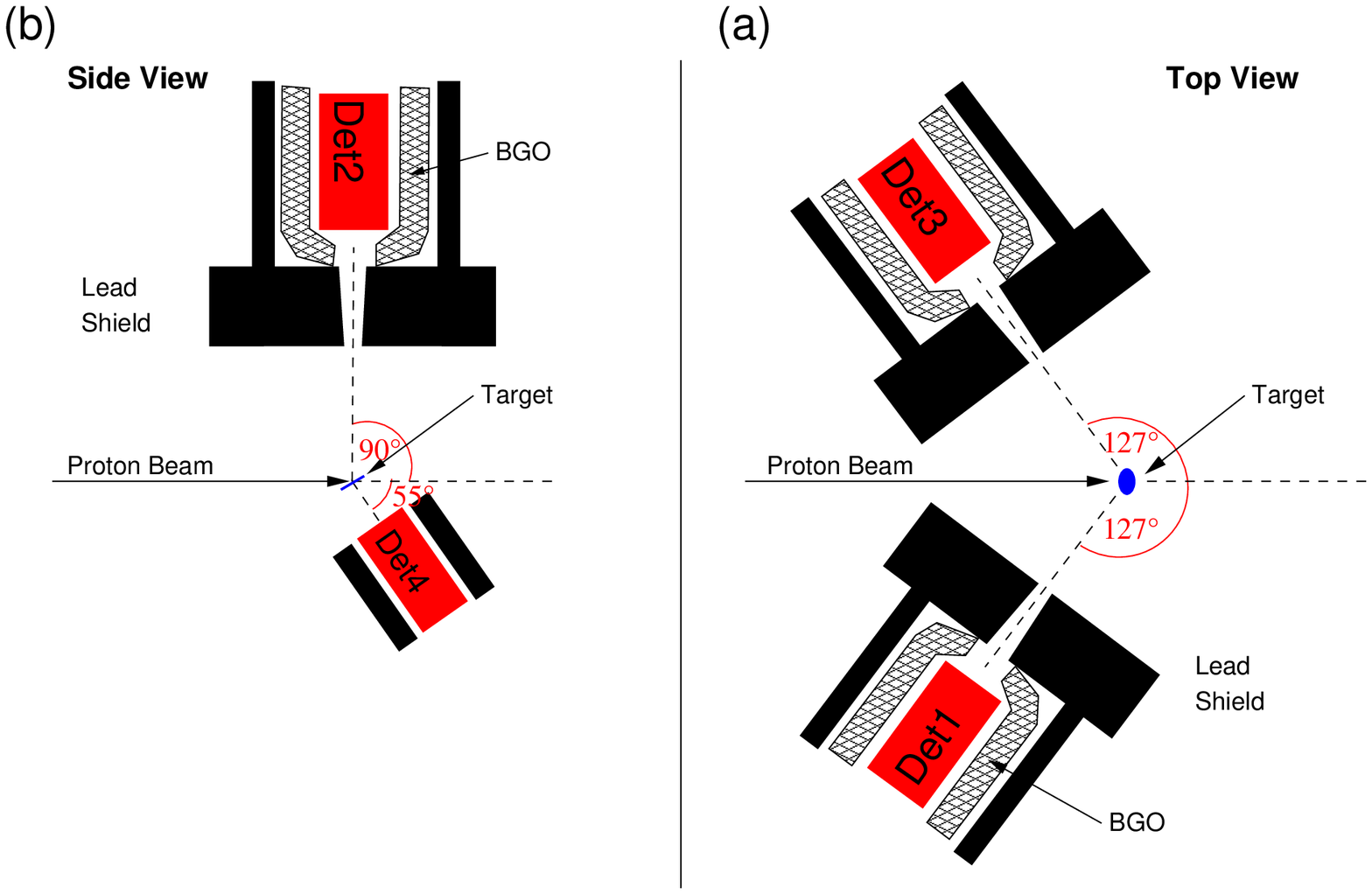}	
	\hspace{5mm}\includegraphics[angle=0,width=0.4\textwidth,trim=0 0 10cm 0,clip]{figure4}	
\caption{(Color online) Scheme of the $\gamma$-ray detection setup. 
Panel (a) shows the two 100\% HPGe detectors (\emph{Det1} and \emph{Det3}) at $\pm$127$^{\circ}$ (left and right).
Panel (b) shows one 100\% HPGe detector (\emph{Det2}) at 90$^{\circ}$ above the target and a 60\% HPGe detector (\emph{Det4}) at 55$^{\circ}$ below the target. 
}
\label{fig:setup}
\end{figure*}

The $\gamma$-ray detection system consisted of four high-purity germanium (HPGe) detectors (fig.~\ref{fig:setup}).  Three 100\% (relative efficiency) HPGe detectors with bismuth germanate (BGO) escape-suppression shield (surrounded by a 2\,cm thick lead shield) and a 10\,cm frontal lead shield with a cone-shaped opening of 3-5\,cm diameter were used: Two were placed horizontally at 127$^{\circ}$ (left and right) relative to the beam direction, with front faces at 32\,cm from the target (hereafter called \emph{Det1} and \emph{Det3}). The third was placed at 90$^{\circ}$ directly above the target, at 28\,cm distance (\emph{Det2}). These three detectors are also used in the nuclear resonance fluorescence (NRF) setup \cite{Schwengner05-NIMA} at the ELBE accelerator. Care was taken so that their shielding and position with respect to the target reproduced the conditions in the NRF setup to $\pm$0.5\,cm.

A fourth smaller HPGe detector (\emph{Det4}, 60\% rel. eff., no escape-suppression, surrounded by a 1\,cm thick lead shield) was placed at 4\,cm distance from the target, at downwards angle 55$^{\circ}$.
This particular setup allowed to observe the emitted photons at three different angles, 55$^{\circ}$, 90$^{\circ}$, and 127$^{\circ}$, and to check the reproducibility for one angle, owing to the two detectors at $\pm$127$^{\circ}$. The second order Legendre polynomial approximately vanishes for angles 55$^{\circ}$ and 127$^{\circ}$, so that angular correlation effects are diluted at these angles. 

\begin{figure}[tbh]
  \includegraphics[angle=0,width=\columnwidth]{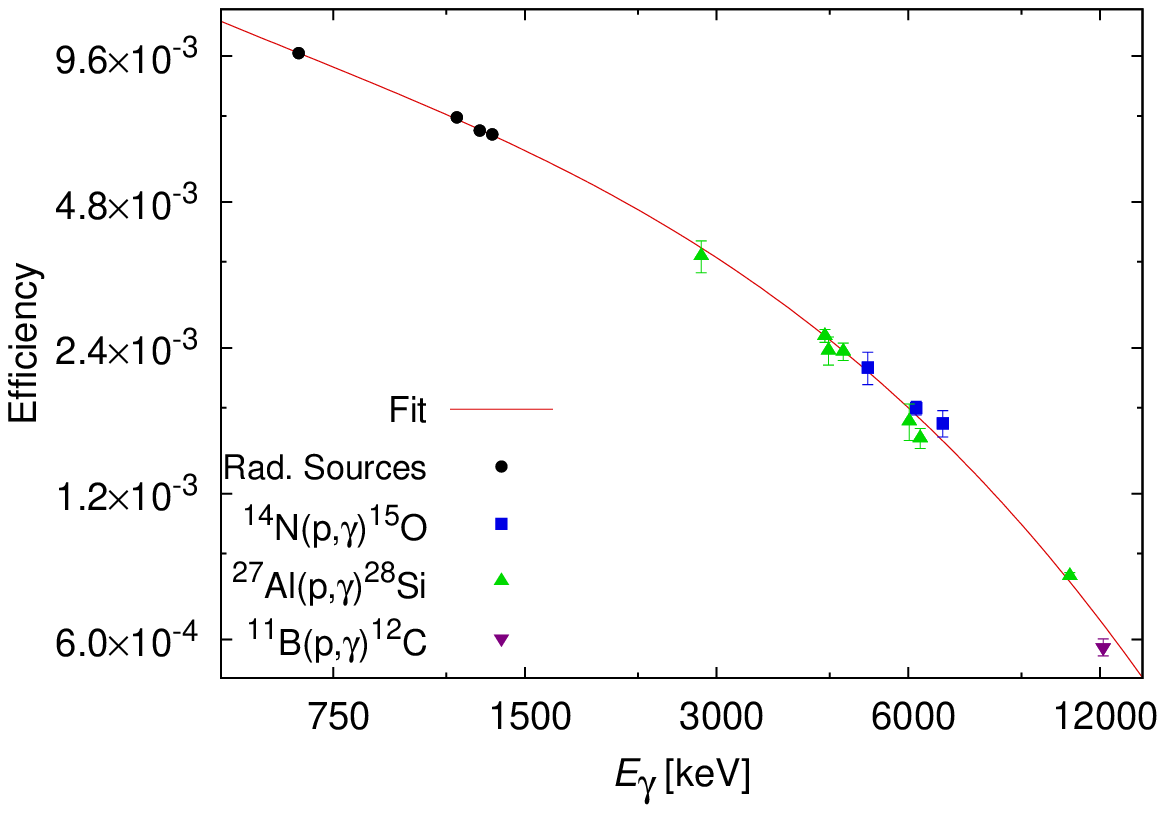} 
\caption{(Color online) Data and parameterization of the full-energy peak $\gamma$-detection efficiency of \emph{Det4}. }
\label{fig:efficiency}
\end{figure}

The $\gamma$-detection efficiencies of the detectors have been measured at low energy (from 662 to 1836\,keV) by means of calibrated radioactive sources ($^{137}$Cs, $^{60}$Co, $^{88}$Y from Physikalisch-Technische Bundesanstalt, quoted 2$\sigma$ relative activity uncertainty 0.8-1.2\%).
The efficiency curve was then extended to higher energy (fig.~\ref{fig:efficiency}) by means of resonant nuclear reaction $\gamma$-cascades of known ratios and angular distributions \cite{Trompler09-Master}. The resonances in $^{11}$B(p,$\gamma$)$^{12}$C at $E_{\rm p}$ = 675\,keV~\cite{Zijderhand90-NIMA}, $^{27}$Al(p,$\gamma$)$^{28}$Si at $E_{\rm p}$ = 992\,keV~\cite{Anttila77-NIM} and $^{14}$N(p,$\gamma$)$^{15}$O at $E_{\rm p}$ = 278\,keV~\cite{Marta08-PRC} were used for this purpose.
For the following analysis, ratios of yields of two high-energy $\gamma$-rays from the same detector have been used. Therefore only $\gamma$-efficiency ratios and not absolute efficiency values were needed.

\section{Experimental procedure}

\subsection{278 and 1058\,keV resonances in $^{14}$N($\rm p$,$\gamma$)$^{15}$O}
\label{subsec:ExpProcedure278+1058}

\begin{figure*}[t]
  \includegraphics[angle=0,width=0.97\textwidth]{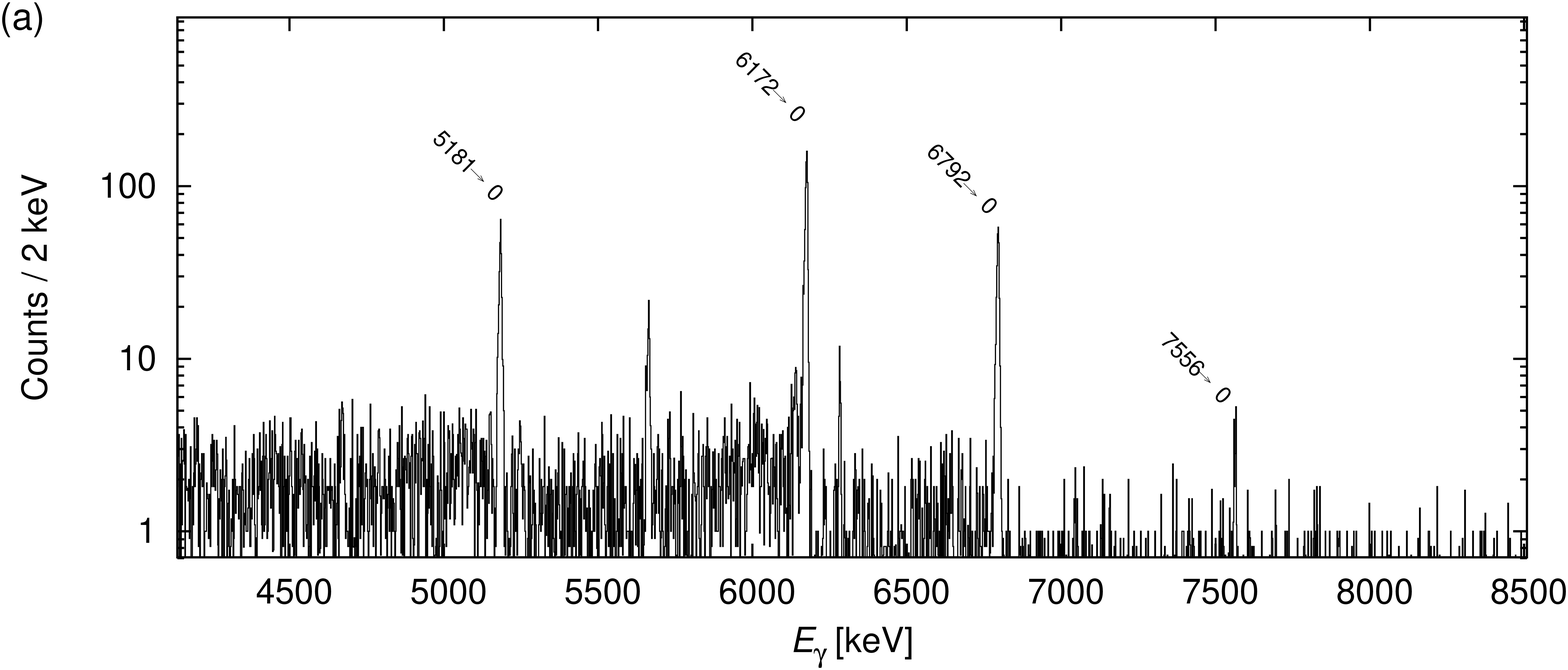} 
  \includegraphics[angle=0,width=1.0\textwidth]{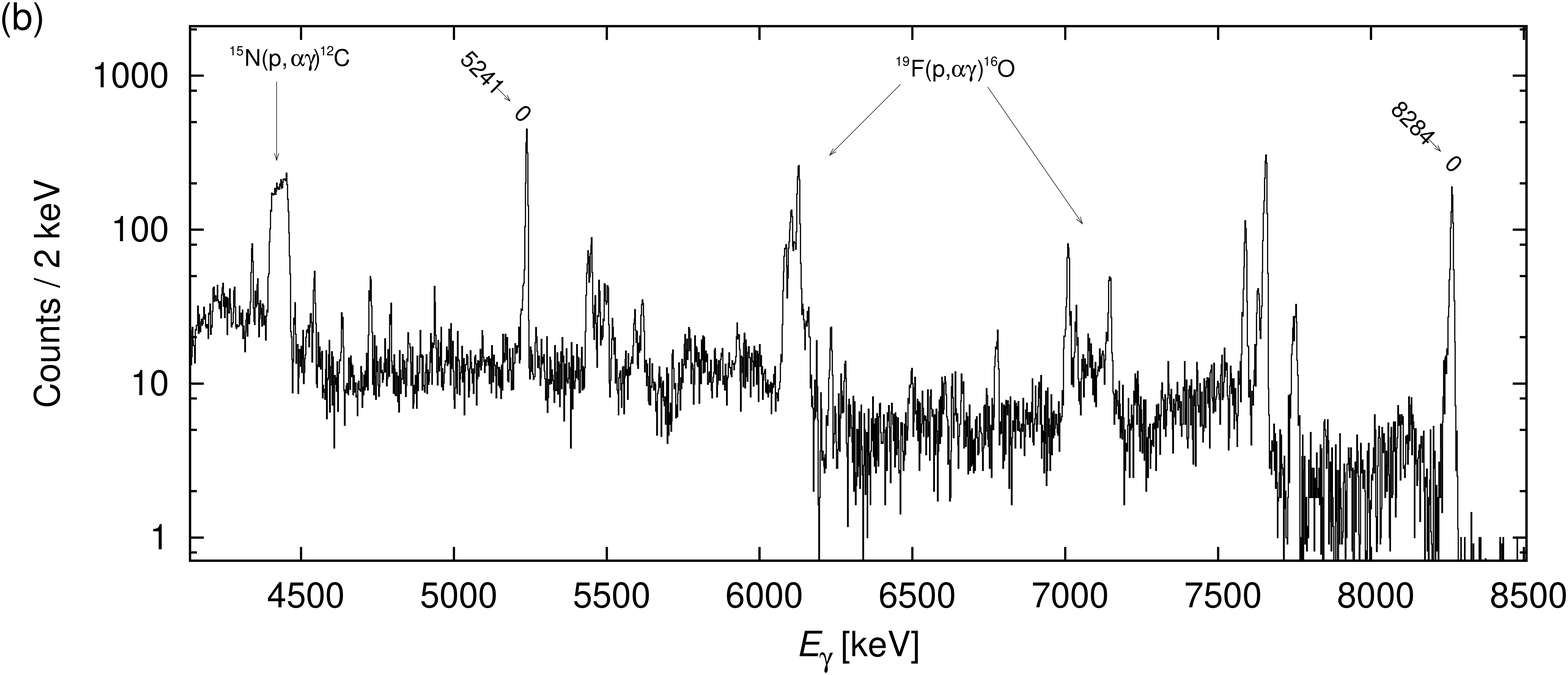} 
\caption{\label{fig:14Nspectra} $\gamma$-ray spectra of \emph{Det1} acquired on the two resonances in the $^{14}$N(p,$\gamma$)$^{15}$O reaction, at $E_{\rm p}$ = 278\,keV (a) and  $E_{\rm p}$ = 1058\,keV (b). The transitions of interest for the reaction under study are marked with tilted tags. The most intense peaks of beam induced background from the $^{19}$F(p,$\alpha\gamma$)$^{16}$O and $^{15}$N(p,$\alpha\gamma$)$^{12}$C reaction are shown as well.} 
\end{figure*}

The $^{14}$N($\rm p$,$\gamma$)$^{15}$O reaction proceeds through radiative capture into one of the states of $^{15}$O (fig.~\ref{fig:levelscheme}, left panel). Non-radiative transitions are negligible. True coincidence summing effects amount to $\leq$ 3\%
($\leq$ 0.5\% uncertainty) in \emph{Det4} and have been corrected for analytically; they are negligible in the other detectors.
Two sharp resonances in the energy range relevant for R-matrix fits have been studied here, at $E_{\rm p}$ = 278 and 1058\,keV (corresponding to $E$ = 259 and 987\,keV, fig.~\ref{fig:levelscheme}, left panel). The proper proton energy for the on-resonance run (fig.~\ref{fig:14Nspectra}) has been chosen based on a scan of the resonance profile, in order to be sure to completely cover its energetic width with the target thickness. 

\begin{figure*}[tbh]
  \includegraphics[angle=-90,width=.49\textwidth]{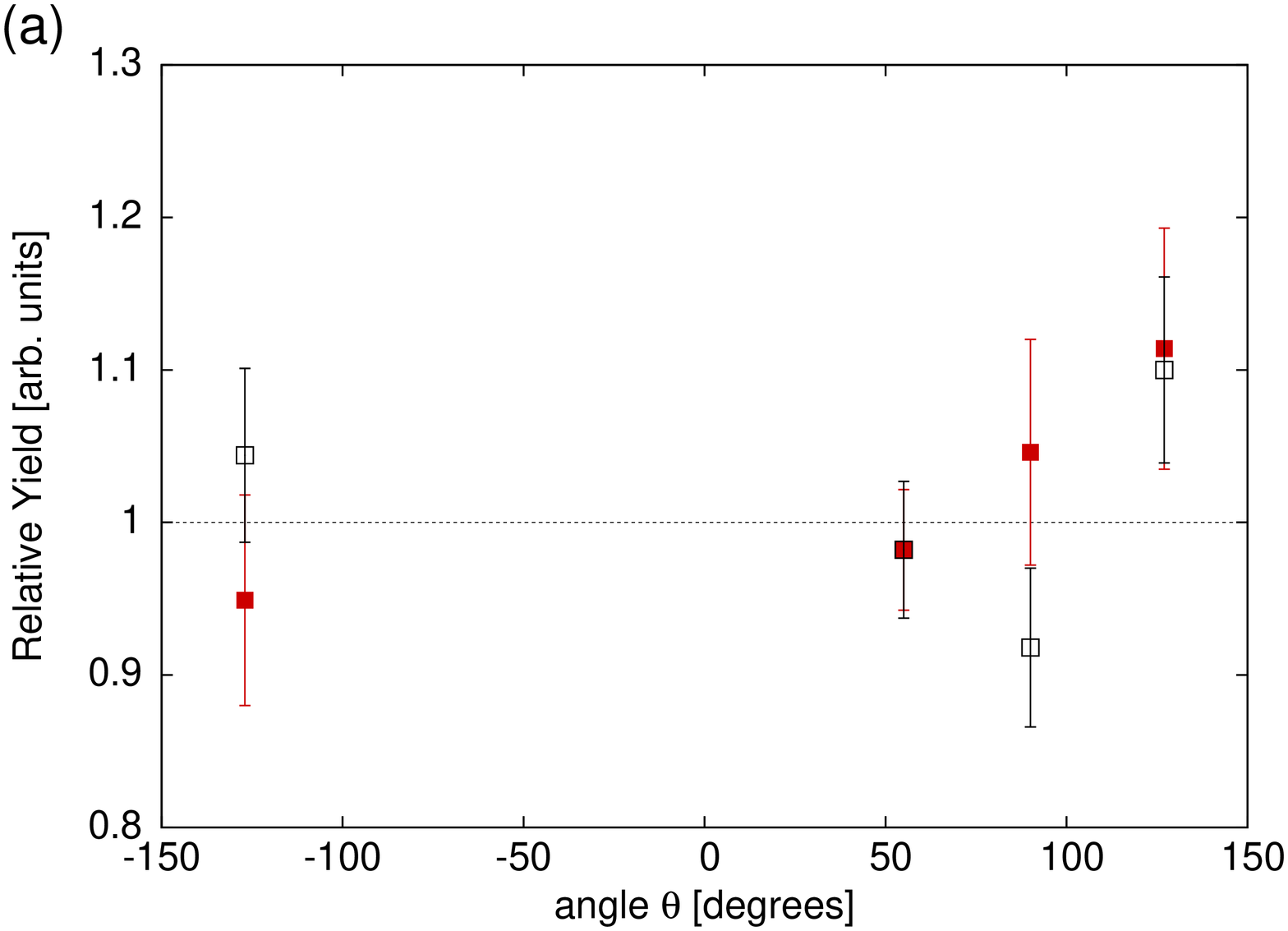}	
  \includegraphics[angle=-90,width=.49\textwidth]{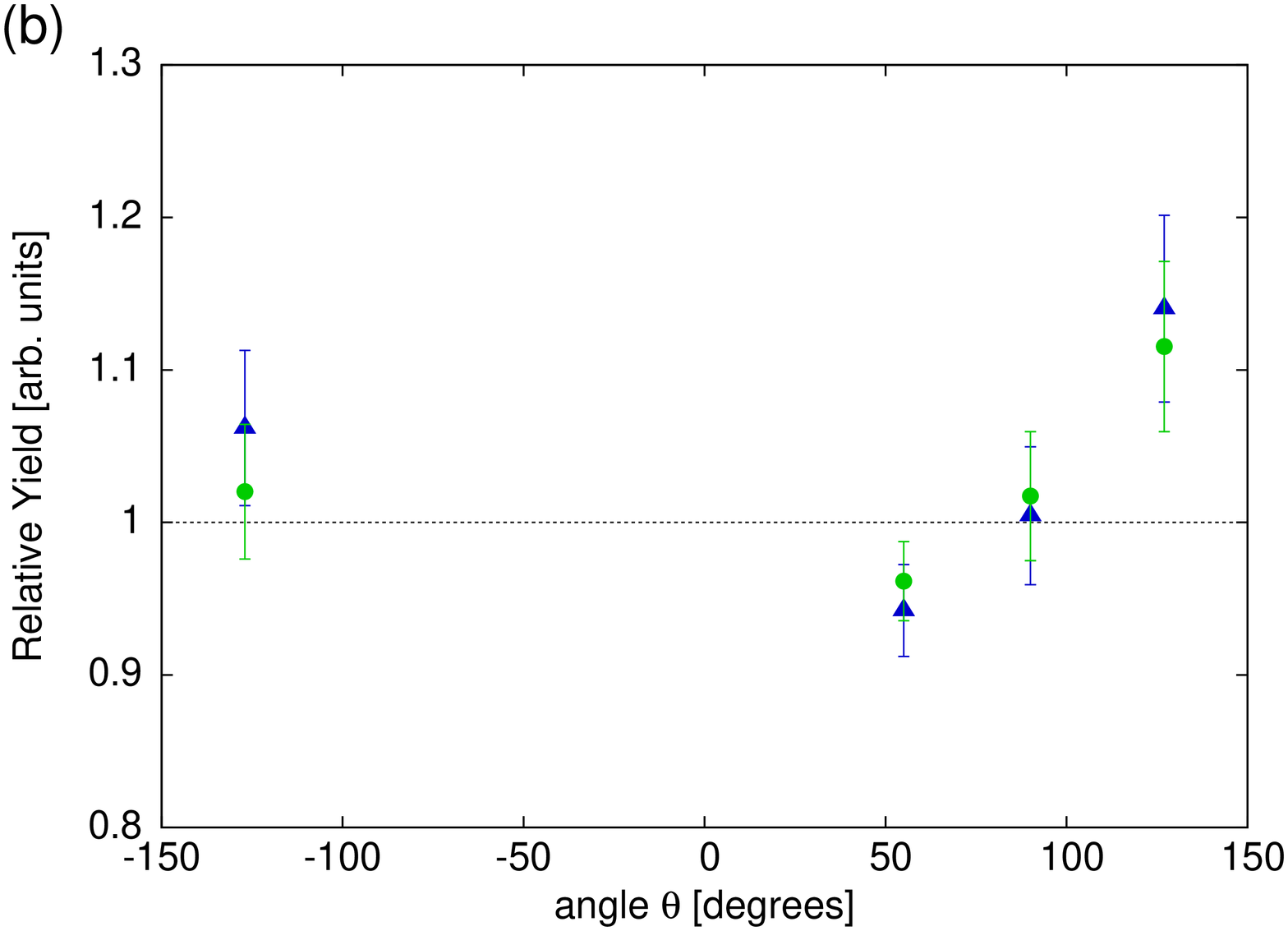}	
\caption{(Color online) (a) Angular distribution of $\gamma$-rays of the $E_{\rm p}$ = 278\,keV resonance in $^{14}$N(p,$\gamma$)$^{15}$O, obtained for 7556$\rightarrow$6172 (red full squares) and 6172$\rightarrow$0 (black empty squares) transitions. (b) The same for the $E_{\rm p}$ = 1058\,keV resonance, 8284$\rightarrow$0 (blue triangles) and 5241$\rightarrow$0 (green circles) transitions.
}
\label{fig:14Nangular}
\end{figure*}

The angular distribution of the 1/2$^+$ resonance at $E_{\rm p}$ = 278\,keV is expected to be isotropic \cite{Ajzenberg13-15,Schroeder87-NPA}. This assumption was experimentally verified here (fig~\ref{fig:14Nangular}, left panel) for transitions through the 6172\,keV state. The present precision is limited by statistics, because the beam intensity of the 3\,MV Tandetron was only 1\,$\mu$A at these low energies. Also the other transitions are found to be isotropic, but within somewhat higher statistical uncertainty. For the present purposes, all $\gamma$-rays from the decay of this resonance are assumed to exhibit isotropy. Combining the data from all four detectors and all transitions, 1.3\% is reached for the statistical uncertainty of the yield of this reference resonance.

For the $E_{\rm p}$ = 1058\,keV resonance, the width was determined here to be $\Gamma_{\rm lab}$ = 3.8$\pm$0.5\,keV, in good agreement with the literature \cite{Ajzenberg13-15}. The proton beam energy chosen for the strength determination was 16\,keV above the resonance energy. Off-resonance runs were performed 
well below and above the resonance, in order to determine and subtract the contribution given by non-resonant capture.
The subtraction amounted to $\approx$ 100\% for the 6792$\rightarrow$0 transition, which proceeds only through the non-resonant mechanism at these energies, and less than 6\% for the 5241$\rightarrow$0 and 8284$\rightarrow$0 transitions.
The angular distribution was checked for the two most intense transitions, i.e. the decay of the 5241\,keV and of the 8284\,keV excited state to the ground state. They were found to be compatible with isotropy within statistics (fig.~\ref{fig:14Nangular}, right panel). For the analysis, isotropy has been assumed and 3\% has been adopted as the uncertainty for the angular distribution.

\subsection{430 and 897\,keV resonances in $^{15}$N($\rm p$,$\alpha\gamma$)$^{12}$C}
\label{subsec:ExpProcedure430+897}

Resonant capture in $^{15}$N(p,$\alpha\gamma$)$^{12}$C proceeds via (1) formation of the compound nucleus $^{16}$O and (2) emission of an $\alpha$ particle 
and a $^{12}$C*(4439) excited nucleus, which then (3) decays to the ground state by emitting a photon (fig.~\ref{fig:levelscheme}, right panel). 
The $E_\gamma$ = 4439\,keV peak is affected by Doppler broadening, with an observed $\gamma$-peak width in \emph{Det4} of 53\,keV for the 430\,keV resonance and 64\,keV for the 897\,keV resonance.

\begin{figure*}[tb]
\centering
 \includegraphics[angle=-90,width=0.49\textwidth]{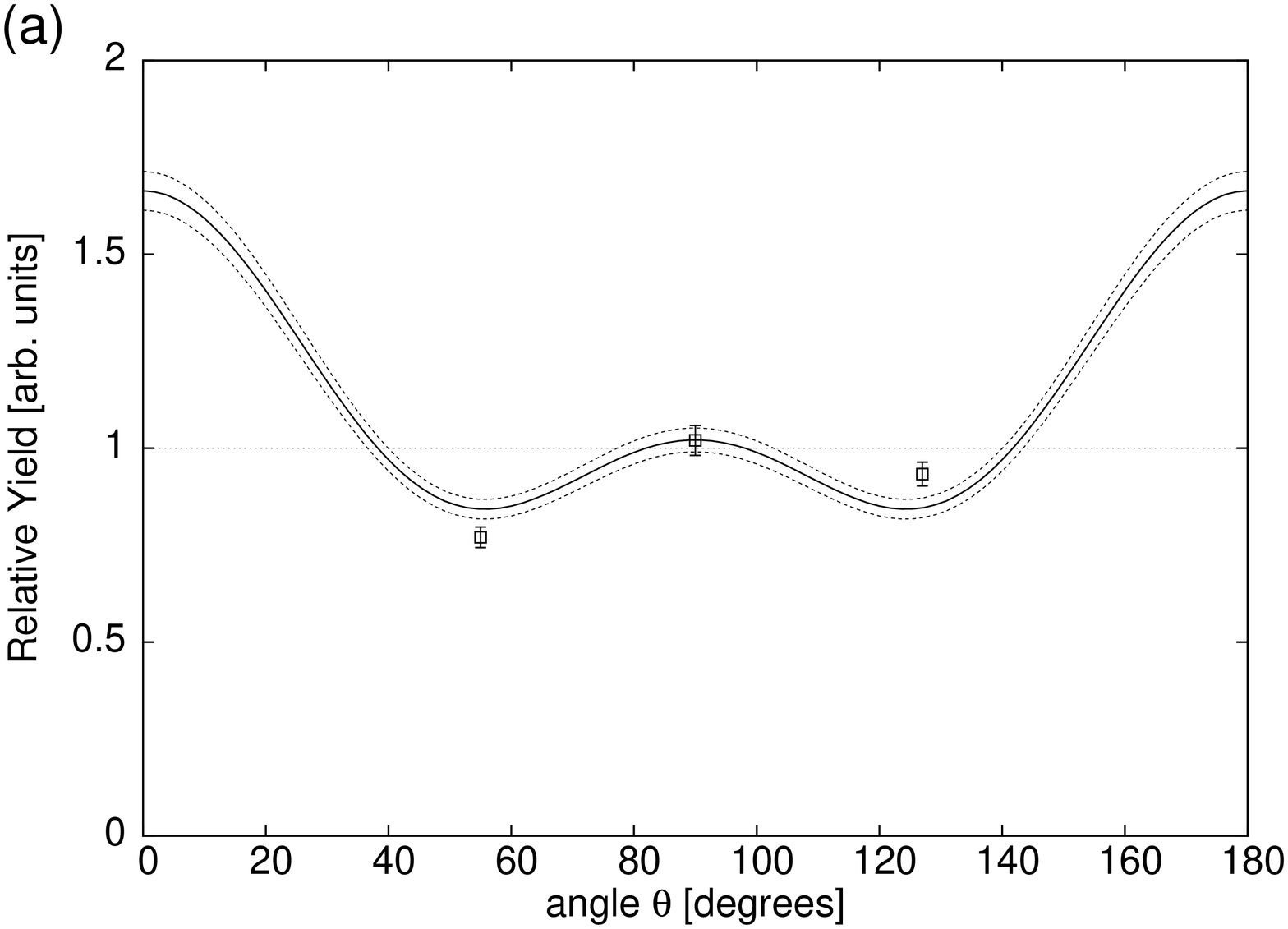}
 \includegraphics[angle=-90,width=0.49\textwidth]{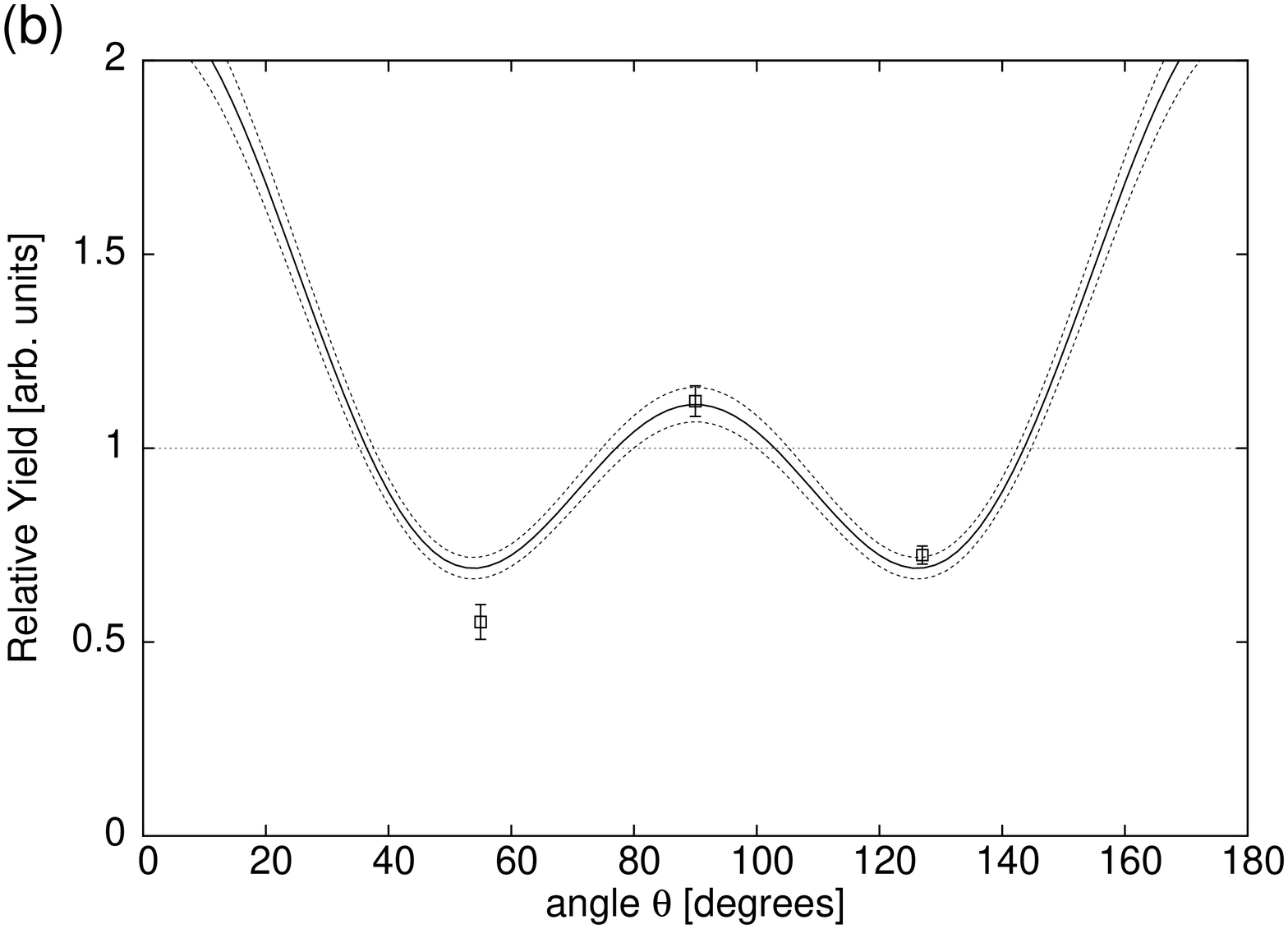}	
 \caption{\label{fig:angular} Angular distribution of $\gamma$-rays emitted by $^{15}$N(p,$\alpha\gamma$)$^{12}$C on the $E_{\rm p}$ = 897\,keV (a) and 430\,keV (b) resonances. The curves parameterize the literature experimental angular distribution \cite{Kraus53-PR}, unfolded 
for attenuation.
The data points are from the present experiment, again unfolded by the respective attenuation coefficients $Q_{2,4}$ (table~\ref{tab:attenuationQ}). The data of \emph{Det1} (-127$^\circ$) and \emph{Det3} (127$^\circ$) are consistent and have been averaged for this plot at 127$^\circ$.
 } 
\end{figure*}

The angular distributions of the 4439\,keV $\gamma$-rays at the two resonances in $^{15}$N(p,$\alpha\gamma$)$^{12}$C are strongly anisotropic but well-known from experiment 
\cite{Kraus53-PR}. The pattern (fig.~\ref{fig:angular}) is similar for both resonances due to the same spin and parity of the excited levels in $^{16}$O and $^{12}$C. The present data are in fair agreement with the literature (fig.~\ref{fig:angular}). For the further analysis, the literature angular distribution has been assumed to be correct. In order to make the angular data comparable, for the close-distance \emph{Det4} non-negligible attenuation coefficients $Q_{2,4}$ calculated based on the prescription given by Ref.~\cite{Rose53-PR} were taken into account (table~\ref{tab:attenuationQ}). These coefficients are consistent with unity for the far-distance detectors \emph{Det1,2,3}. 

\begin{table*}[tb]
 \caption{\label{tab:attenuationQ} 
Experimental yield ratio 430/897 of the two $^{15}$N(p,$\alpha\gamma$)$^{12}$C resonances at different angles: Uncorrected value (column 8), and value that was corrected for angular effects (column 9). The attenuation factors $Q_{2,4}$ calculated following Ref.~\cite{Rose53-PR} and the literature angular distribution coefficients $W_i(\theta)$ \cite{Kraus53-PR} are also given (columns 4-5 and 6-7, respectively). 
 } 
\centering

\begin{tabular}{|lrr|r@{$\pm$}lr@{$\pm$}l|r@{$\pm$}lr@{$\pm$}l|r@{$\pm$}lr@{$\pm$}l|}
\hline
 & & & \multicolumn{4}{c|}{Calculated} & \multicolumn{4}{c|}{Literature \cite{Kraus53-PR}} & \multicolumn{2}{c}{Uncorrected}	 & \multicolumn{2}{c|}{Corrected}	\\
Detector & \multicolumn{1}{c}{$\theta$} & d\,[cm] & \multicolumn{2}{c}{$Q_2$} & \multicolumn{2}{c|}{$Q_4$}	& \multicolumn{2}{c}{$W_{430}(\theta)$}	& \multicolumn{2}{c|}{$W_{897}(\theta)$} & \multicolumn{4}{c|}{exp. yield ratio 430/897} \\[1mm]
\hline
\emph{Det1} [present] 	& -127$^\circ$ &	32 & \multicolumn{2}{c}{$1.00^{+0.00}_{-0.01}$}	& \multicolumn{2}{c|}{$1.00^{+0.00}_{-0.01}$}	&	0.69 & 0.01	 & 0.85 & 0.01	 & 0.071 & 0.002	 & 0.087 & 0.004	 \\
\emph{Det2} [present]	& 90$^\circ$ & 28 &	\multicolumn{2}{c}{$1.00^{+0.00}_{-0.01}$}	& \multicolumn{2}{c|}{$1.00^{+0.00}_{-0.01}$}	&		1.11 & 0.02	 & 1.02 & 0.02	 & 0.100 & 0.003	 & 0.091 & 0.004	 \\
\emph{Det3} [present] 	& 127$^\circ$ &	32 &	\multicolumn{2}{c}{$1.00^{+0.00}_{-0.01}$}	& \multicolumn{2}{c|}{$1.00^{+0.00}_{-0.01}$}	&	0.69 & 0.01	 & 0.85 & 0.01	 & 0.070 & 0.002	 & 0.086 & 0.004	 \\
\emph{Det4} [present]	& 55$^\circ$ & 4 &	0.70 & 0.05	& 0.25 & 0.09	&		0.92 & 0.03	 & 0.96 & 0.01	 & 0.076 & 0.001	 & 0.079 & 0.003	 \\
\hline
\emph{Det4'} \cite{Caciolli10-PhD}	& 0$^\circ$ &	10 &	0.94 & 0.02	& 0.80 & 0.04	& 1.98 & 0.05	 & 1.57 & 0.05	 & 0.106 & 0.001	& 0.084 & 0.004	\\
\hline

\end{tabular}

\end{table*}

As a reliability check, the ratio 430/897 of the yields of the 4439\,keV $\gamma$-peak for two consecutive runs on the two different resonances was calculated for all detectors (table~\ref{tab:attenuationQ}). The same ratio has also been calculated for a similar experiment \cite{Caciolli10-PhD} with targets enriched in $^{15}$N and \emph{Det4'} placed at 0$^\circ$, 
where the anisotropy is very pronounced, and 10\,cm distance (table~\ref{tab:attenuationQ}, last line). The yield ratio depends only on the effective detection angle of the device, hence the angular distribution and its attenuation. After correcting for these two effects, the values for the yield ratio are consistent (table~\ref{tab:attenuationQ}).

\section{Data analysis and Results}
\label{sec:results}

\subsection{Branching ratios for the decay of 1058\,keV resonance in $^{14}$N(p,$\gamma$)$^{15}$O}
\label{subsec:results_branching}

The branching ratios for the decay of the $E_{\rm p}$ = 1058\,keV resonance have been measured using the high-statistics spectra of \emph{Det4} (table~\ref{tab:14Nbranching}), with the off-resonant contribution subtracted based on reference runs below and above the resonance. Since \emph{Det4} is located at 55$^\circ$ where the second order Legendre polynomial vanishes, angular corrections have been neglected for all transitions. For the two strongest transitions, this assumption was verified experimentally (sec.~\ref{subsec:ExpProcedure278+1058}). The branching ratios were determined also for some of the weaker transitions. The branching ratios in the standard compilation \cite{Ajzenberg13-15} are based on one work \cite{Evans66-PR}. The only exception is the weak 8284$\rightarrow$5181 branch reported by Ref.~\cite{Schroeder87-NPA}, which was adopted, leading to a recalculation of the other branches \cite{Ajzenberg13-15}.

\begin{table*}[tb]
\caption{Branching ratios, in \%, for the decay of the resonance at $E_{\rm p}$ = 1058\,keV in $^{14}$N(p,$\gamma$)$^{15}$O from the literature \cite{Evans66-PR,Schroeder87-NPA}, from the standard compilation \cite{Ajzenberg13-15} based on these papers, and from the present work. 
See text for a discussion of the new recommended values.}
\centering
	\begin{tabular}{lr@{$\pm$}lr@{$\pm$}lr@{$\pm$}lr@{$\pm$}lr@{$\pm$}l}
	\hline
	$E_{\rm x}$ [keV]	& \mc{2}{r}{Ref.~\cite{Evans66-PR}} & \mc{2}{r}{Ref.~\cite{Schroeder87-NPA}} & \mc{2}{r}{Compilation \cite{Ajzenberg13-15}} & \mc{2}{r}{Present work} & \mc{2}{l}{New recommended}	\\		\hline
 6859 &  1.2 & 0.3\fnm{1}  &  1.2 & 0.2      &  1.2 & 0.3      &  1.2 & 0.3        &  \phantom{\hspace{1cm}}1.2 & 0.2\\
 6172 &  2.2 & 0.6\fnm{1}  &  2.6 & 0.1      &  2.2 & 0.6      &  1.7 & 0.4        &  2.5  & 0.1\\
 5241 & 42.7 & 0.5\fnm{1}  & 46   & 2        & 42.2 & 0.5      & 45.4 & 0.8        & 45.2  & 0.7\\
 5181 & \mc{2}{c}{}        &  1.2&0.1\fnm{1} &  1.2 & 0.1      &  1.1 &  0.6       &  1.2 & 0.1 \\
    0 & 53.8 & 0.25\fnm{1} & 49   & 2        & 53.2 & 0.25      & 50.5 &  0.8       & 49.9  & 0.7 \\
	\hline
	\end{tabular}
\footnotetext[1]{Adopted in the compilation \cite{Ajzenberg13-15}.}
\label{tab:14Nbranching} 
\end{table*}

For the two strongest transitions, 8284$\rightarrow$0 and 8284$\rightarrow$5241, the present branchings are in agreement with Ref.~\cite{Schroeder87-NPA}, but not with Ref.~\cite{Evans66-PR}. The present data show the 8284$\rightarrow$5241 transition to be stronger than reported in Ref.~\cite{Evans66-PR}. In that work \cite{Evans66-PR}, a sodium iodide scintillating detector had been used that was surrounded with a large Compton-suppressing guard detector. It is conceivable that the guard efficiency correction applied in Ref.~\cite{Evans66-PR} might have been different for the single 8284$\rightarrow$0 $\gamma$-ray than for the $\gamma$-rays of the 8284$\rightarrow$5241$\rightarrow$0 cascade, leading to some systematic uncertainty. The present values for the weaker transitions 8284$\rightarrow$6859, 8284$\rightarrow$6172, and 8284$\rightarrow$5181 are in good agreement with the literature \cite{Evans66-PR,Schroeder87-NPA} but show generally lesser precision. 

Due to the significant differences observed in the strongest two branches, new recommended values are necessary for future calibration purposes. For 8284$\rightarrow$0 and 8284$\rightarrow$5241, the outlying values by Ref.~\cite{Evans66-PR} are omitted and a weighted average of Ref.~\cite{Schroeder87-NPA} and the present data is formed. For the other three transitions, a weighted average of Refs.~\cite{Evans66-PR, Schroeder87-NPA} and the present data is adopted (table~\ref{tab:14Nbranching}).

\subsection{Relative resonance strengths}
\label{subsec:results_strength}

The total width of the three resonances under study here is small compared to the energy loss in the present targets (table~\ref{tab:allReson}). Therefore, the classical definition of the thick target yield \cite{Iliadis07-Book} is applicable:
\begin{equation}
\label{eq:infYield}
Y_{\infty}^i = \frac{\lambda^2}{2} \, \beta^i \, \frac{\omega\gamma}{\epsilon} \, ; 
\hspace{1.5cm}
Y_{\infty} = \frac{\sum_i Y_{\infty}^i}{\sum_i \beta^i} = \frac{\lambda^2}{2} \, \frac{\omega\gamma}{\epsilon}
\end{equation}
where $Y_{\infty}^i$ is the experimental yield for branch $i$ with branching ratio $\beta^i$ corrected for $\gamma$-efficiency and angular distribution, and $\lambda$ is the de Broglie wavelength at the resonance energy. $\epsilon$ is the effective stopping power \cite{Iliadis07-Book}, i.e. the stopping power per nucleus taking part in the reaction under study. If the target of interest is $^{14}$N, $\epsilon$ is given by:
\begin{equation}
\epsilon^{14}(E_{\rm p})=\epsilon_{\rm N}(E_{\rm p})(1+\frac{n_{^{15}\rm N}}{n_{^{14}\rm N}})+\epsilon_{\rm Ti}(E_{\rm p})\frac{n_{\rm Ti}}{n_{^{14}\rm N}} 
\end{equation}
and analogously for $^{15}$N as target:
\begin{equation}
\label{eq:effDe}
 \epsilon^{15}(E_{\rm p})=\epsilon_{\rm N}(E_{\rm p})(1+\frac{n_{^{14}\rm N}}{n_{^{15}\rm N}})+\epsilon_{\rm Ti}(E_{\rm p})\frac{n_{\rm Ti}}{n_{^{15}\rm N}} = \frac{n_{^{14}\rm N}}{n_{^{15}\rm N}} \epsilon^{14}(E_{\rm p}).
\end{equation}
The isotopic abundance $n_{^{15}\rm N}$/$n_{^{14}\rm N}$ is always taken to be the standard value, 0.3663
/99.6337 \cite{Coplen02-PAC}, with an uncertainty of 1.0\% \cite{Junk58-GCA}. 
The ratio of resonance strengths for two different resonances at $E_{\rm p}$ = $n$\,keV ($n \in \{430;897;1058\}$) and at $E_{\rm p}$ = 278\,keV, the reference strength, is then given by:
\begin{equation}
\label{eq:Rstrength}
 \frac{\omega\gamma_n}{\omega\gamma_{278}} = \frac{Y_{\infty,n}}{Y_{\infty,278}}\, \frac{\lambda_{278}^2}{\lambda_{n}^2} \, \frac{\epsilon^{a}(n)}{\epsilon^{14}(278)}; \qquad a \in \{14;15\}.	
\end{equation}
The ratio of yields $Y_{\infty,n}$/$Y_{\infty,278}$ was taken from the weighted average of the ratios obtained for each of the four detectors, after checking that they were consistent. 
The ratio of effective stopping powers at different energies $\epsilon^{a}(n)$/$\epsilon^{14}(278)$ is only slightly dependent on the target stoichiometry $n_{\rm Ti}/n_{^{14}\rm N}$. The main uncertainty associated with stopping powers is their absolute scale and not the energy dependence beyond the Bragg peak \cite{Ziegler08-Book}, and only the energy dependence is needed here. The stoichiometric ratio varied for the worst case from Ti$_1$N$_{0.93}$ (virgin target) to Ti$_1$N$_{0.80}$ (after a H$^+$ dose of 0.97\,Coulomb). Using the stopping powers from {\tt SRIM} \cite{SRIM08.02}, this change affected $\epsilon^{14}(\rm 1058)$/$\epsilon^{14}(278)$ by just 0.1\%. In order to include also theoretical uncertainties, 1.0\% uncertainty is assumed for $\epsilon^{a}(n)$/$\epsilon^{14}(278)$.	

The target deterioration under beam bombardment has been corrected for based on the change observed in the yield of the $E_{\rm p}$ = 897\,keV resonance in $^{15}$N(p,$\alpha\gamma$)$^{12}$C that was used for the regular target scans (fig.~\ref{fig:target}), leading to 0.9\% uncertainty. 

For calculating the reference yield of the $E_{\rm p}$ = 278\,keV resonance, the yields of the three peaks corresponding to the decay of the $E_{\rm x}$ = 6792, 6172, and 5182\,keV excited states of $^{15}$O and their precisely known branching ratios \cite{Imbriani05-EPJA} have been used.
The strength of the $E_{\rm p}$ = 1058\,keV resonance has been obtained based on the yields from the two strongest transitions, 5241$\rightarrow$0 and 8284$\rightarrow$0, and the presently measured branching ratios (sec.~\ref{subsec:results_branching}, tab.~\ref{tab:14Nbranching}).

For the two resonances in $^{15}$N(p,$\alpha\gamma$)$^{12}$C, the broad $\gamma$-peak at 4439\,keV was used to calculate the yield.
Their strength ratio was found to be $\omega\gamma_{430}/\omega\gamma_{897}$ = (6.25$\pm$0.17)$\cdot$10$^{-2}$, in fair agreement with the literature value of (5.8$\pm$0.2)$\cdot$10$^{-2}$. That value had been obtained with two detectors placed at 55$^\circ$ \cite{Zijderhand86-NPA}, neglecting angular distribution effects and the resultant uncertainty. The present error bar includes these effects. Because of a target change, the ratio $\omega\gamma_{430}/\omega\gamma_{278}$ had to be calculated in two steps
\begin{equation}
\frac{\omega\gamma_{430}}{\omega\gamma_{278}} = \frac{\omega\gamma_{430}}{\omega\gamma_{897}} \frac{\omega\gamma_{897}}{\omega\gamma_{278}} 
\end{equation}
leading to slightly higher uncertainty. All the errors for the resonance strength ratios are summarized in table~\ref{tab:errors}.

Using these strength ratios and the reference strength $\omega\gamma_{278}$ = 13.1$\pm$0.6\,meV \cite{Seattle09-workshop}, new absolute resonance strengths have been obtained for the three resonances under study (table~\ref{tab:allReson}). 

\begin{table*}[tb]
\caption{\label{tab:allReson} Relative and absolute resonance strength values $\omega\gamma$. The errors for the new absolute $\omega\gamma$ values include the uncertainty from the reference strength $\omega\gamma_{278}$. See the text for a discussion of the new recommended values. }
\centering
\begin{tabular}{|l|r|c|r@{$\pm$}l|r@{$\pm$}l|r@{$\pm$}l|r@{$\pm$}l|}
\hline
Reaction 	& \multicolumn{2}{c|}{Literature \cite{Ajzenberg13-15,Tilley_16-17}} & \multicolumn{4}{c|}{Present} & \multicolumn{2}{c|}{Literature} & \multicolumn{2}{c|}{New recommended}	\\
 	& $E_{\rm p}$ [keV]	& $\Gamma_{\rm lab}$ [keV] & \multicolumn{2}{c|}{$\omega\gamma_n/\omega\gamma_{278}$} & \multicolumn{2}{c|}{$\omega\gamma$ [eV]} & \multicolumn{2}{c|}{$\omega\gamma$ [eV]} & \multicolumn{2}{c|}{$\omega\gamma$ [eV]}	\\
\hline
$^{14}$N(p,$\gamma$)$^{15}$O	& 278  & 1.12\fnm{1} & \multicolumn{2}{c|}{$\stackrel{\rm Def}{=}$1}	 & \multicolumn{2}{c|}{---} &	0.0131 & 0.0006\fnm{2} & \multicolumn{2}{c|}{---}\\ 
$^{14}$N(p,$\gamma$)$^{15}$O	& 1058 & 3.8\fnm{3} &	27.8 & 0.9  &	\phantom{00} 0.364 & 0.021	&	0.31 & 0.04~\cite{Schroeder87-NPA} &	 \phantom{00} 0.353 & 0.018 \\
$^{15}$N(p,$\alpha\gamma$)$^{12}$C	& 430  & 0.1 &	(1.73 & 0.07)$\cdot$10$^3$  &	22.7 & 1.4	&	21.1 & 1.4~\cite{Becker95-ZPA} &	 21.9 & 1.0 \\	%
$^{15}$N(p,$\alpha\gamma$)$^{12}$C \phantom{0}	& 897  & 1.57 &	(2.77 & 0.09)$\cdot$10$^4$  &	362 & 20	 &	293 & 38~\cite{Zijderhand86-NPA} &	362 & 20 \\
\hline
\end{tabular}
\footnotetext[1]{Ref.~\cite{Borowski09-AIPCP}.}
\footnotetext[2]{Weighted average \cite{Seattle09-workshop} of Refs.~\cite{Becker82-ZPA,Imbriani05-EPJA,Runkle05-PRL,Lemut06-PLB}.}
\footnotetext[3]{Literature: 3.9$\pm$0.7\,keV \cite{Ajzenberg13-15}. Present work: 3.8$\pm$0.5\,keV (sec.\,\ref{subsec:ExpProcedure278+1058}).}
\end{table*}

\begin{table}[tb]
\caption{\label{tab:errors} Uncertainties affecting the resonance strength ratios $\omega\gamma_n/\omega\gamma_{\rm ref}$. The 1058 and 897\,keV resonances are referred to the 278\,keV resonance, $\omega\gamma_{\rm ref} = \omega\gamma_{278}$. The 429\,keV resonance is referred to the 897\,keV resonance, $\omega\gamma_{\rm ref} = \omega\gamma_{897}$. Its uncertainty includes also the 1.0\% from the isotopic abundance $n_{^{15}\rm N}$/$n_{^{14}\rm N}$.
}
\centering
\begin{tabular}{|lccc|}
\hline
	 &	1058\,keV	 &	897\,keV	 &	430\,keV	 \\
\hline
Counting statistics	 &	\textbf{1.7\%}	 &	\textbf{1.6\%}	 &	\textbf{1.0\%}	 \\
$\gamma$-efficiency (relative) \cite{Trompler09-Master}	 &	0.7\%	 &	1.3\%	 &	 	 \\
Decay branching ratio	 &	1.2\%	 &	0.5\%	 &	 	 \\
Angular distribution	 &	1.8\%	 &	1.1\%	 &	1.9\%	 \\
Stopping power ratio &	1.0\%	 &	1.0\%	 &	1.0\%	 \\
Isotopic abundance $n_{^{15}\rm N}$/$n_{^{14}\rm N}$	 &		 &	1.0\%	 &	 	 \\
Target degradation $n_{^{14}\rm N}$/$n_{\rm Ti}$	 &	0.9\%	 &	0.9\%	 &	0.8\%	 \\
Beam intensity &	1.0\%	 &	1.0\%	 &	1.0\%	 \\
\textbf{Final uncertainty of $\omega\gamma_n/\omega\gamma_{\rm ref}$}	 &	\textbf{3.3\%}	 &	\textbf{3.1\%}	 &	\textbf{2.7\%}	 \\
\hline
Reference strength $\omega\gamma_{\rm ref}$	 &	4.6\%	 &	4.6\%	 &	5.5\%	 \\
\textbf{Final uncertainty of $\omega\gamma_n$}	 &	\textbf{5.7\%}	 &	\textbf{5.5\%}	 &	\textbf{6.1\%}	 \\
\hline
\end{tabular}
\end{table}

\section{Discussion}

Near the 1058\,keV resonance in $^{14}$N(p,$\gamma$)$^{15}$O, R-matrix fits for the strongest contribution, ground state capture, show a pronounced interference pattern \cite{Angulo01-NPA,Mukhamedzhanov03-PRC,Marta08-PRC}. Therefore, the shape of the excitation curve for this transition does not obey the ideal Breit-Wigner form. 
Since the present, rather thick target covers the entirety of the energy range directly affected by the resonance, the present strength value is unaffected by this fact.
Still, it should be noted that due to the interference, the formal R-matrix parameters for this resonance are quite far from the experimental values. 
The present and more precise strength value can therefore not be used directly in an R-matrix code. However, in the future it can be compared with the predicted strength from an updated R-matrix code with the proper resonance treatment \cite[e.g.]{Azuma10-PRC}, as soon as such a code is publicly available.

For the other branches of the 1058\,keV resonance and also for all the other resonances under study here, such an interference pattern either does not exist or is negligible when compared to the on-resonance capture. 

The present strength value of the 1058\,keV resonance in $^{14}$N(p,$\gamma$)$^{15}$O is higher than the previous number \cite{Schroeder87-NPA}, but still in agreement within the uncertainty. Therefore, a weighted average of the two numbers is formed and recommended for future use (table~\ref{tab:allReson}). 

Also for the 897\,keV resonance in $^{15}$N(p,$\alpha\gamma$)$^{12}$C, the present value is higher than the literature \cite{Zijderhand86-NPA}. That value \cite{Zijderhand86-NPA} had been obtained just with two detectors at 55$^\circ$ angle and neglecting angular distribution effects. However, the literature angular distribution \cite{Kraus53-PR} is lower than unity at 55$^\circ$ (fig.~\ref{fig:angular}, also confirmed by the present data) so this assumption leads to a systematically low value. Consequently, the $\omega\gamma$ value from the present experiment is recommended for future use. 

For the 430\,keV resonance, the present strength, determined based on $\gamma$-spectroscopy, has the same precision as the literature value which had been obtained by $\alpha$-spectroscopy instead \cite{Becker95-ZPA}. That work \cite{Becker95-ZPA} had used an $\alpha$-detector at 30$^\circ$ and applied the $\alpha$-particle angular distribution from a previous experiment and R-matrix fit \cite{Leavitt83-NPA}.
Based on the two independent results from $\alpha$-spectroscopy \cite{Becker95-ZPA} and from $\gamma$-spectroscopy (present work), a weighted average for the strength is recommended that has just 4\% uncertainty (table~\ref{tab:allReson}).

\section{Summary and outlook}
\label{sec:summary}

The resonance strength $\omega\gamma$ has been measured for the 1058\,keV resonance  in $^{14}$N(p,$\gamma$)$^{15}$O and the 430 and 897\,keV resonances in $^{15}$N(p,$\alpha\gamma$)$^{12}$C, relative to the well-known strength of the 278\,keV resonance in $^{14}$N(p,$\gamma$)$^{15}$O. A called-for improvement in the precision of this reference point \cite{Seattle09-workshop} will therefore also lead to an improvement in the understanding of the other three resonances.

For the major transitions, the angular distributions of the 278 and 1058\,keV resonances in $^{14}$N(p,$\gamma$)$^{15}$O have been verified experimentally to be consistent with the expected isotropy. The decay branching ratios of the 1058\,keV resonance in $^{14}$N(p,$\gamma$)$^{15}$O have been determined and updated values are recommended. 

Three well-understood, sharp resonances are now available as natural normalization points for cross section measurements. 
The new, precise strength of the 430\,keV resonance in $^{15}$N(p,$\alpha\gamma$)$^{12}$C has the potential to serve as a highly precise standard value to make hydrogen depth profiling absolute.
The road is paved for a re-measurement of the astrophysically important $^{14}$N(p,$\gamma$)$^{15}$O off-resonance cross section at energies near 1\,MeV.

\begin{acknowledgments}
The support by the staff and operators of the FZD ion beam center, technical support by Michael Fauth, Andreas Hartmann, and Manfred Sobiella (FZD), and target analyses performed by Alberto Vomiero (CNR Brescia, Italy) are gratefully acknowledged. 
This work was supported in part by the European Union, Research Infrastructures Transnational Access (RITA 025646) to the AIM facility, by DFG (BE4100/2-1), and by OTKA (T68801). T.S. acknowledges support from the Herbert Quandt Foundation.
\end{acknowledgments}


\end{document}